\documentclass[english,aps,a4paper,prd,superscriptaddress,nofootinbib,preprintnumbers,preprint,floatfix,showpacs,12pt]{revtex4}
\usepackage[T1]{fontenc}
\usepackage{graphicx}
\usepackage{amssymb, latexsym}
\usepackage{slashed}
\usepackage{multirow}
\usepackage{booktabs}
\usepackage{amsmath}
\usepackage{wasysym}

\usepackage{mathrsfs}   %\mathscr{L}
\usepackage{slashed}     %\slashed{p}
\usepackage{url} \usepackage{graphicx}
\usepackage[colorlinks=true,linkcolor=redLinks,citecolor=greenLinks,urlcolor=redLinks,
pdfborder={0 0 1}]{hyperref} 
\usepackage{xcolor} 
\usepackage{framed}
\usepackage{natbib} 
\definecolor{greenLinks}{rgb}{0, 0.6, 0} 
\definecolor{blueLinks}{rgb}{0, 0, 0.6}
\definecolor{redLinks}{rgb}{0.8, 0, 0}
\definecolor{tempText}{rgb}{0.55, 0.10,0.67}
\definecolor{eprintLinks}{rgb}{0.4, 0.4, 0.4}
\definecolor{journalLinks}{rgb}{0.6, 0, 0}

\newcommand{\MYhref}[3][redLinks]{\href{#2}{\color{#1}{#3}}}%

\catcode`\@=11
\def\lsim{\mathrel{\mathpalette\@versim<}}
\def\gsim{\mathrel{\mathpalette\@versim>}}
\def\@versim#1#2{\vcenter{\offinterlineskip
\ialign{$\m@th#1\hfil##\hfil$\crcr#2\crcr\sim\crcr } }}
\catcode`\@=12

\def\lnv{lepton number violation }
\def\lnc{lepton number conservation }

\usepackage{float}
\newcommand{\bea}{\begin{eqnarray}}
\newcommand{\eea}{\end{eqnarray}}
\def\SM{$\mathrm{SU(3)_c \otimes SU(2)_L \otimes U(1)_Y}$ }
\newcommand{\sm}{Standard Model }
 
 \def\z53{ $\mathbb{Z}_{5} \otimes \mathbb{Z}_{3} $}
\def\ze33{ $\mathbb{Z}^D_{3} \otimes \mathbb{Z}_{3} $}
\newcommand{\AddrUNAM}{ {\it Instituto de F\'{\i}sica, Universidad Nacional Aut\'onoma de M\'exico, A.P. 20-364, Ciudad de M\'exico 01000, M\'exico.}}
\newcommand{\AddrAHEP}{
  {\it AHEP Group, Instituto de F\'{\i}sica Corpuscular --
    C.S.I.C./Universitat de Val{\`e}ncia \\
    Edificio de Institutos de Paterna,
 C/Catedratico Jos\'e Beltr\'an, 2 E-46980 Paterna (Val\`{e}ncia) - SPAIN}}

\begin{document}
\title{Two--loop Dirac neutrino mass and WIMP dark matter}
\author{Cesar Bonilla}
\email{cesar.bonilla@ific.uv.es}\affiliation{\AddrAHEP} \author{Ernest
  Ma} \email{ma@phyun8.ucr.edu} \affiliation{Physics Department,
  University of California, Riverside, California 92521, USA}
\author{Eduardo Peinado}
\email{epeinado@fisica.unam.mx}\affiliation{\AddrUNAM}
\author{Jose W.F. Valle}
\email{valle@ific.uv.es}\affiliation{\AddrAHEP}


\begin{abstract}

  We propose a ``scotogenic'' mechanism relating small neutrino mass
  and cosmological dark matter. Neutrinos are Dirac fermions with
  masses arising only in two--loop order through the sector
  responsible for dark matter.  Two triality symmetries ensure both
  dark matter stability and strict \lnc at higher orders. A global
  spontaneously broken U(1) symmetry leads to a physical $Diracon$
  that induces invisible Higgs decays which add up to the Higgs to
  dark matter mode. This enhances sensitivities to spin--independent
  WIMP dark matter search below $m_h/2$.

\end{abstract}
\maketitle
\section{Introduction}

Two of the main observational shortcomings of the \sm is that it lacks
neutrino masses~\cite{Valle:2015pba} as well as a viable candidate for
cosmological dark matter~\cite{Bertone:2004pz}. Even though light
neutrinos themselves can account only for a very small fraction of the
dark matter, they may hold the key to the basic understanding of what
causes the dark matter to exist in the first place.
Indeed, the existence of neutrino masses and of cosmological dark
matter may be closely interconnected in several
ways~\cite{Valle:2012yx}.
For example, the mechanism of neutrino mass generation itself can
involve the exchange of particles which make up the bulk of the
observed dark matter.
This is the main idea of scotogenic models~\cite{Ma:2006km,Kubo:2006yx}.
The prototype model is based on the assumption that the dark sector,
odd under a parity symmetry, is connected with the neutrino sector
through the generation of the light neutrino masses. The dark matter
particle plays the role of messenger of radiative neutrino mass
generation~\cite{Hirsch:2013ola,Merle:2016scw}.
In the simplest conventional scenario~\cite{Ma:2006km}, the dark
matter is made up of a weakly interacting massive particle (WIMP), for
example, the lightest scalar component of an inert Higgs doublet.

In this letter we explore the possibility of generating scotogenic
Dirac neutrino masses radiatively, by forbidding \lnv through the
cyclic $Z_3$ symmetry. This ensures strict \lnc and the Diracness of
neutrinos at higher orders.
Another discrete $Z_3$ symmetry is responsible for dark matter
stability. Neutrino masses are generated only at the two--loop
level, through the same sector responsible for cosmological dark
matter.
This new realization combines the idea of two--loop scotogenic
neutrino masses suggested in~\cite{Ma:2007gq} with the idea of having
a conserved lepton number leading to the Dirac nature of neutrinos.
In our model there is a global U(1) symmetry which forbids the usual
Dirac mass term of the neutrinos with the standard
Higgs~\cite{Bento:1991bc,Peltoniemi:1992ss}~\footnote{In contrast to
  Refs.~\cite{Bento:1991bc,Peltoniemi:1992ss}, neutrinos here are
  Dirac fermions, as opposed to
  Quasi-Dirac~\cite{valle:1982yw}.}. This symmetry breaks
spontaneously leading to a physical Goldstone boson -- a gauge singlet
$Diracon$~\cite{Bonilla:2016zef} -- which induces invisible Higgs
decays. These are analogous to the invisible Higgs decays by Majoron
emission in models with Majorana
neutrinos~\cite{joshipura:1992hp}. The extra invisible channel adds up
to the Higgs boson decays to pairs of dark matter particles at
collider experiments, providing tighter limits on WIMP dark matter
below $m_h/2$.

\section{The model}
\label{sec:Models}

We will consider a simple extension of the standard \SM model with the
symmetries and field content indicated in Table~\ref{tab:scotochido}, 
where $\omega$ and $\alpha$ are cube roots of unity,
i.e. $\omega^3=1=\alpha^3$.
%% .
There are two complex $SU(2)_L$ doublets, $H$ and $\eta$ and three
complex singlets, $\sigma$, $\chi$ and $\xi$.
\begin{table}[h!]
\begin{center}
\begin{tabular}{|c||c|c|c||c|c|c||c|c|c|c|}
\hline
& $\bar{L}$ & $\nu^c$ & $H$& $\eta$ & $N$ &  $S$& $\sigma$ &  $\xi$ &  $\chi$\\
\hline\hline
$SU(2)_L$ & 2 & 1 & 2 & 2 & 1 & 1 & 1 & 1 &1\\
\hline
$U(1)_{D}$ & $-1$ & $3$ & $0$ & $0$ & $-1$ & $1$& $2$ & $-2$ & $0$  \\
\hline
$Z_3^{DM}$ & $1$ & $1$ & $1$  & $\alpha$ & $\alpha$ & $\alpha$& $1$ & $\alpha^2$ & $\alpha$ \\
\hline
$Z_3$ & $\omega$ & $\omega^2$ & $1$  & $1$ & $\omega$ & $\omega^2$& $1$ & $1$ & $1$ \\
\hline
\end{tabular}\caption{Relevant particle content and quantum numbers of the 
  model. } \label{tab:scotochido}
\end{center}
\end{table}

The invariant Yukawa Lagrangian is given by
\begin{equation}
{\cal L}= y^\nu \bar{L}\tilde{\eta} S+y^R\nu^c N \xi+\lambda N S \chi.
\end{equation}
The scalar potential can be separated as follows
\begin{equation}
{\cal V}=V+V(H,\eta)+V(\xi,\sigma,\chi,H,\eta),
\end{equation}
where the first term $V$ in the scalar potential contains the
relevant terms for the generation of the neutrino masses, namely,
\begin{equation}
V = \lambda^\chi_1  H^{\dagger} \eta\chi^*+\lambda^\chi_2 \chi  \sigma \xi +\lambda^\chi_3 \chi^3+ h.c.
\end{equation}
while the second term $V(H,\eta)$ is the Higgs potential associated to
the $\eta$ doublet
\begin{equation}
\begin{array}{lcl}
V(H,\eta)&=&\mu_1^2 H^\dagger H+\mu_2^2 \eta^\dagger \eta+ \lambda_1 \vert H \vert^4+\lambda_2  \vert \eta \vert^4+\lambda_3 \vert H\vert^2 \vert \eta \vert^2 \\ \\&+& \lambda_4 \vert H^\dagger \eta\vert^2 + {\lambda_5} \left[(H^\dagger \eta)^2+ h.c.\right].
\end{array}
\end{equation} 
The last term $V(\xi,\sigma,\chi,H,\eta)$, is given by
\begin{equation}\begin{array}{lcl}
V(\xi,\sigma,\chi,H,\eta)&=&\mu_\xi^2 \xi \xi^* +\mu_\sigma^2 \sigma \sigma^*+ \mu_\chi^2 \chi\chi^*+\lambda_\xi  (\xi \xi^*)^2+\lambda_\sigma (\sigma \sigma^*)^2+\lambda_\chi  (\chi\chi^*)^2+\lambda_{\sigma\xi} \sigma \sigma^*\xi \xi^*\\
\\&+&\lambda_{\chi\xi}\chi \chi^*\xi \xi^*+\lambda_{\chi\sigma}\chi \chi^* \sigma \sigma^*+\lambda_{\chi H}\chi \chi^*H^\dagger H+\lambda_{\chi \eta}\chi \chi^*\eta^\dagger \eta
+\lambda_{\xi H}\xi\xi^*H^\dagger H\\ \\&+&\lambda_{\xi \eta}\xi\xi^*\eta^\dagger \eta+\lambda_{\sigma H}\sigma\sigma^*H^\dagger H+\lambda_{\sigma\eta}\sigma\sigma^*\eta^\dagger \eta.
\end{array}
\end{equation}
After spontaneous symmetry breaking the fields are shifted as follows,
\begin{equation}\begin{array}{ccc}
H=\left(\begin{array}{c}H^{+}\\\frac{1}{\sqrt{2}}(v_h+h^0+iA^0)\end{array}\right),& &\eta=\left(\begin{array}{c}\eta^{+}\\\frac{1}{\sqrt{2}}(\eta_R+i\eta_I)\end{array}\right),\\ \\
\sigma=\frac{1}{\sqrt{2}}(v_\sigma+\sigma_R+i\sigma_I),&\chi=\frac{1}{\sqrt{2}}(\chi_R+i\chi_I),&\xi=\frac{1}{\sqrt{2}}(\xi_R+i\xi_I).
\end{array}\end{equation}
Notice that there are no vacuum expectation values (vevs)
for scalars $\eta, \chi, \xi$ which are charged under $Z_3^{DM}$.
\begin{figure}[!h]
\centering
\includegraphics[scale=0.5]{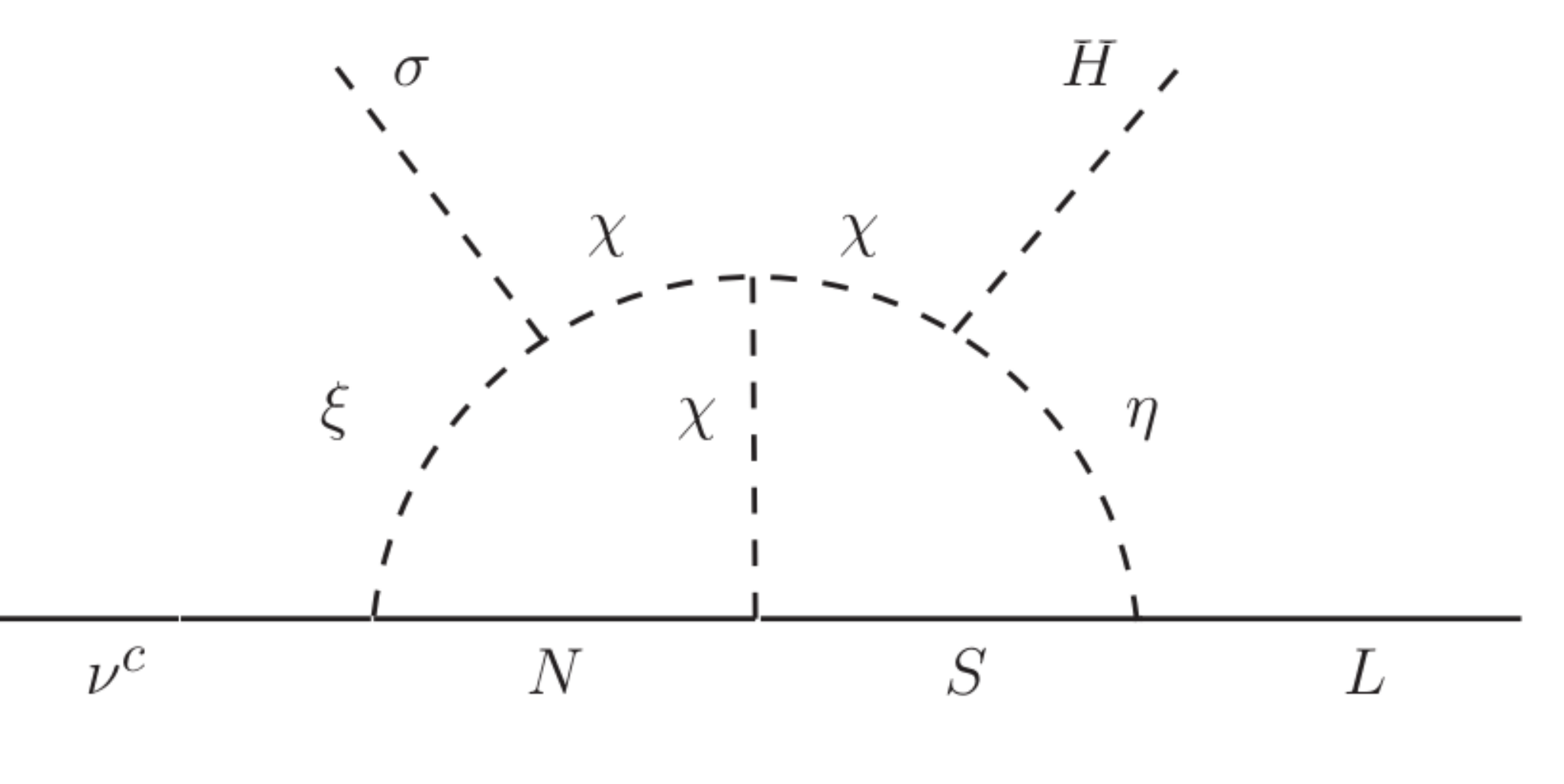} 
\caption{Two--loop generation of Dirac neutrino mass.}
\label{fig:loop}
\end{figure}

The fermions $N$ and $S$ will form heavy Dirac neutrinos by pairing
with their corresponding partners $\bar{N}$ and $\bar{S}$. The light
neutrinos acquire their masses via the loop in Fig.~\ref{fig:loop}.
From the minimization of the scalar potential, the scalar fields
charged under the $Z_3^{DM}$ do not acquire vacuum expectation values
(vevs), while for the Higgs and the $\sigma$ fields, there are two
relevant tadpole equations:
\begin{equation}\begin{array}{l}
\mu_1^2=\lambda_1 v_h^2+\frac{1}{2}\lambda_{\sigma H}v_\sigma^2,\\
\mu_\sigma^2=\lambda_2 v_\sigma^2+\frac{1}{2}\lambda_{\sigma H}v_h^2.\\
\end{array}
\end{equation}

The corresponding mass matrix for the CP-even ``active'' scalars is
\begin{equation}
M_R^2=\left(
\begin{array}{cc}
 2 \lambda_{1} v_h^2  &\lambda_{\sigma H} v_h v_\sigma \\
\lambda_{\sigma H} v_h v_\sigma & 2\lambda_\sigma v_\sigma^2
\end{array}
\right).
\end{equation}

The pseudoscalars include the unphysical Goldstone boson $G^0$ and a
physical one, $\mathcal{D}$, namely the $Diracon$. In contrast to that
of Ref.~\cite{Bonilla:2016zef} the $Diracon$ here is a pure singlet
under weak SU(2) and hence is not subject to the strong astrophysical
bound coming from stellar cooling considerations~\cite{raffelt1996stars}.

The mass matrices for the scalar and pseudoscalar ``dark'' sector
(charged under $Z_3^{DM}$) in the basis $\eta, \chi, \xi$ are given as
\begin{equation}
\label{inertR}
\mathcal{M}_R^{2}=\left(
\begin{array}{ccc}
 \frac{1}{2} \left(\lambda_{345}^+ v_h^2+ \lambda_{\sigma \eta}v_\sigma^2-2\mu_\eta^2\right) & \frac{\lambda^\chi_1v_h}{2 \sqrt{2}} & 0 \\
 \frac{\lambda^\chi_1v_h }{2 \sqrt{2}} & \frac{1}{2} \left(\lambda_{\chi H}v_h^2+\lambda_{\xi \sigma}v_\sigma^2 \right) & \frac{\lambda^\chi_2v_\sigma}{\sqrt{2}} \\
 0 & \frac{\lambda^\chi_2v_\sigma}{\sqrt{2}}& \frac{1}{2} \left(\lambda_{\xi H}v_h^2+\lambda_{\sigma \xi}v_\sigma^2  \right) 
\end{array}
\right),
\end{equation}
and
\begin{equation}
\label{inertI}
\mathcal{M}_I^{2}=\left(
\begin{array}{ccc}
 \frac{1}{2} \left(\lambda_{345}^- v_h^2+ \lambda_{\sigma \eta}v_\sigma^2-2\mu_\eta^2\right) &- \frac{\lambda^\chi_1v_h}{2 \sqrt{2}} & 0 \\
 -\frac{\lambda^\chi_1v_h }{2 \sqrt{2}} & \frac{1}{2} \left(\lambda_{\chi H}v_h^2+\lambda_{\xi \sigma}v_\sigma^2 \right) & -\frac{\lambda^\chi_2v_\sigma}{\sqrt{2}} \\
 0 &- \frac{\lambda^\chi_2v_\sigma}{\sqrt{2}}& \frac{1}{2} \left(\lambda_{\xi H}v_h^2+\lambda_{\sigma \xi}v_\sigma^2  \right)
\end{array}
\right),
\end{equation}
where the parameter $\lambda^{\pm}_{345}$ is given by
\begin{equation}\label{lam345}\lambda^{\pm}_{345}\equiv\lambda_3+\lambda_4\pm\lambda_5.\end{equation}
Finally, the mass for the ``inert'' electrically charged scalar is given by
\begin{equation}
\mathcal{M}_{\eta^+}^2=\frac{1}{2} \left(-2 \mu_\eta^2+\lambda_3 v_h^2+\lambda_{\sigma\eta} v_\sigma^2\right).
\end{equation}

\section{ Dark matter annihilation}
\label{sec:ModRes}

As usual in scotogenic
models~\cite{Ma:2006km}~\cite{Hirsch:2013ola,Merle:2016scw} dark
matter in our model can be either scalar or fermionic.
Here we focus on the first case, where the dark matter candidate is
the lightest scalar eigenstate of $\mathcal{M}^{2}_{R,I}$ in
Eqs.~(\ref{inertR}) and (\ref{inertI}), which can be a general mixture
of $Z_3^{DM}$--charged doublet and singlet scalars $\eta, \chi$ and
$\xi$.
An important requirement for the dark matter interpretation of
such candidate is that its relic abundance matches the value
observed by the Planck collaboration.
There are in principle three possibilities~\footnote{In order to
  generate nonzero neutrino mass through Fig.~\ref{fig:loop} none of
  the $\lambda^{\chi}_i$ couplings can vanish exactly. Hence the dark
  matter candidate is necessarily a combination of the
  triality--carrying scalars.}:
\begin{itemize}
\item mainly doublet dark matter
\item generic doublet--singlet dark matter combination
\item mainly singlet dark matter
\end{itemize}
The first case can be arranged if the coupling $\lambda^{\chi}_1$ is
suppressed and/or the vev of $\sigma$ is large.
In this case one looses the signature corresponding to invisible Higgs
decay to the $Diracon$, Eq.~(\ref{eq:inv}).
The dark matter candidate is well studied in other similar scenarios
such as the scotogenic model \cite{Ma:2006km} and the Inert Doublet
Model~\cite{Deshpande:1977rw,Cao:2007rm,Ilnicka:2015jba}. In this
context, the sign of the dimensionless coupling $\lambda_5$,
determines whether the dark matter has a either CP-odd or CP-even
nature, and the correct relic abundance constrains the parameter
$\lambda_{345}^{\pm}$ in Eq.~(\ref{lam345})~\cite{Ilnicka:2015jba}.

In the second and most general case the situation is analogous to that
of sneutrino dark matter in the inverse seesaw model described in
Ref.~\cite{arina:2008bb,An:2011uq}. The dark matter candidate is made up of a
singlet-doublet combination with potentially ``comparable''
components, and can lead both to an adequate relic density as well as
to a detectable signal in nuclear recoil.

Finally, the last and simplest of the three cases, corresponds to that
in which the dark matter candidate is mainly singlet and is detected
primarily by the Higgs portal interaction.
In the present model, the dark matter singlet would be given mainly by
a combination of the fields $\chi$ and $\xi$. Without loss of
generality we will denote as $X$ the lightest combination of these
singlets~\footnote{We assume that the doublet-singlet mixing is
  negligible.  Then, we define
  $X\equiv c_{\alpha} \chi - s_{\alpha} \xi$ and
  $\bar{X}\equiv s_{\alpha} \chi + c_{\alpha} \xi$.}.

However, thanks to the $Z_3^{DM}$ nature of our dark matter candidate
and to the presence of the $Diracon$, there are other distinctive
features in our case.
Indeed, due to the cubic terms in the scalar potential, one finds
that, besides annihilations, semi-annihilation processes play an
important role in determining the dark matter relic density, as
explained carefully in Ref.~\cite{Belanger:2014bga}. In contrast to
the case of $Z_2$ dark matter, the dark matter spin--independent
direct detection cross section is no longer directly related to the
annihilation cross section.

In the case of interest, the limit in which the dark matter candidate
$X$ is stabilized by the $Z_3^{DM}$ symmetry has been studied in
detail in Ref.~\cite{Belanger:2012zr,Belanger:2014bga}.  In this case
the dimensionful term $\lambda_\chi^3$ contributes to the
semi-annihilation processes like, for instance, $XX\to X^{\ast} h$
that can dominate in the determination of the relic density.
As a result the $\lambda_{XH}$ coupling no longer links the
annihilation rate to the spin independent nuclear recoil detection
cross section, in contrast to the more familiar case in which dark
matter is stabilized by the $Z_2$ symmetry~\cite{Feng:2014vea}.

\begin{figure}[!h]
\centering
  \includegraphics[width=0.3\textwidth]{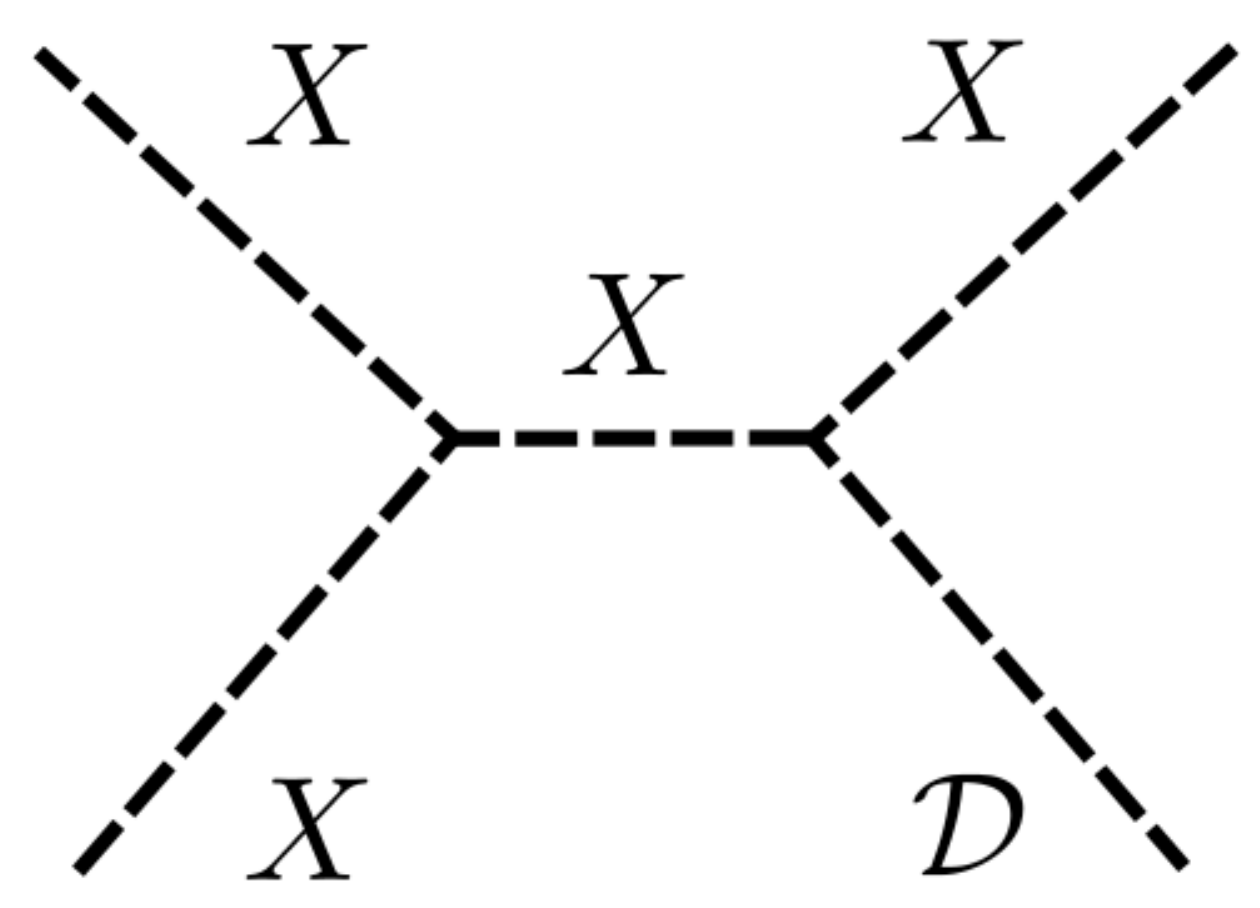} \ \
  \includegraphics[width=0.3\textwidth]{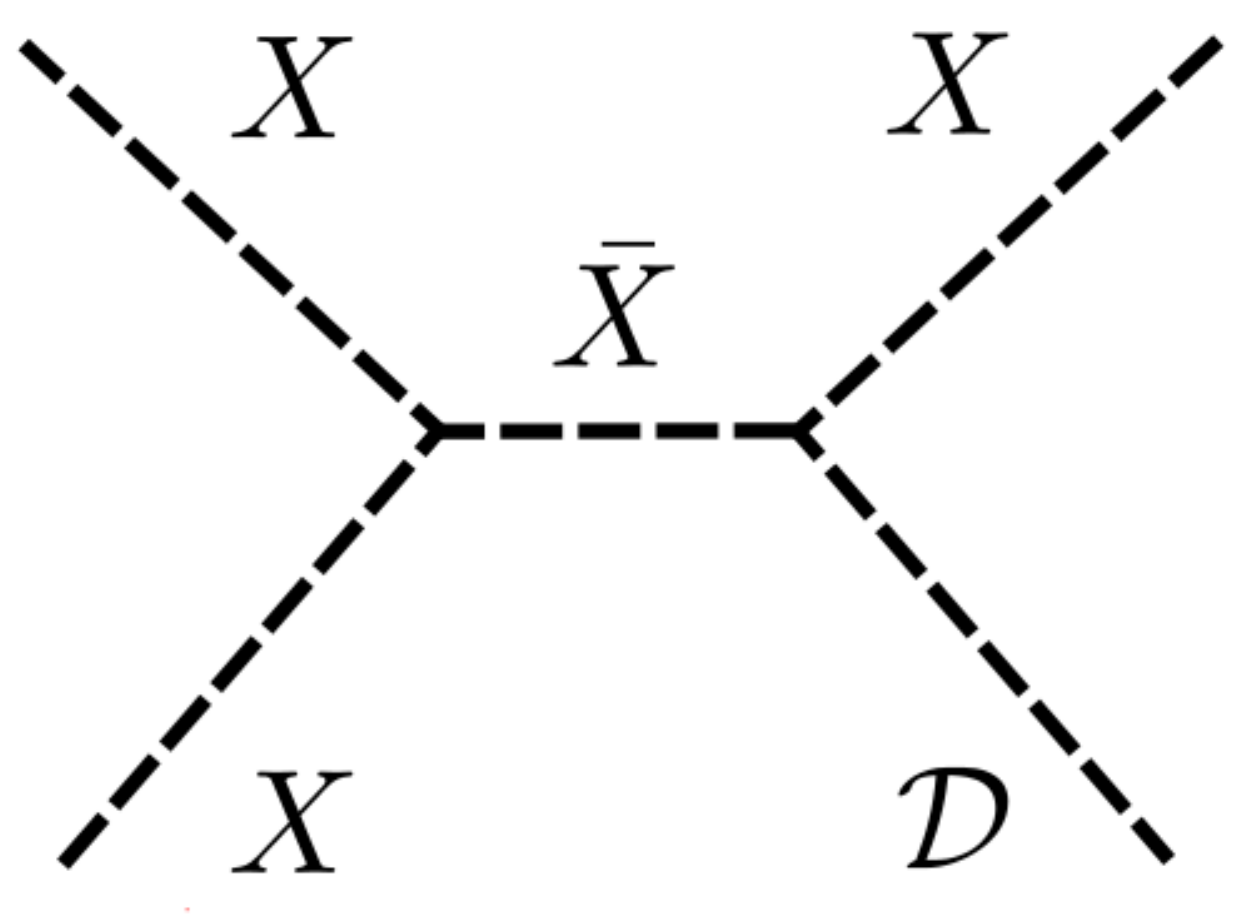} \\
  \caption{Dark matter semi-annihilation channels involving the
    $Diracon$.}
\label{fig:fey}
\end{figure}
Over and above this observation, our model has further distinctively
novel features associated to the presence of the $Diracon$.
This leads to genuinely new interactions absent in previous dark
matter models, including the simplest benchmark model studied
in~\cite{Feng:2014vea} as well as the possibilities analyzed in
Refs.~\cite{Belanger:2012zr,Belanger:2014bga}. Indeed, concerning dark
matter annihilation, there are new semi-annihilation channels
involving the $Diracons$, as illustrated in
Fig.~\ref{fig:fey}~\footnote{A detailed determination of the relic
  density lies outside the scope of this paper. }.
These which should allow one to suppress the $X$ relic density with
respect to the cases considered in these references.
In addition, the $Diracon$ plays a role in detection, see next.

\section{Dark matter detection}
\label{sec:dark-matt-detect}

\begin{figure}[h!]
\centering
\includegraphics[height=6cm,width=8cm]{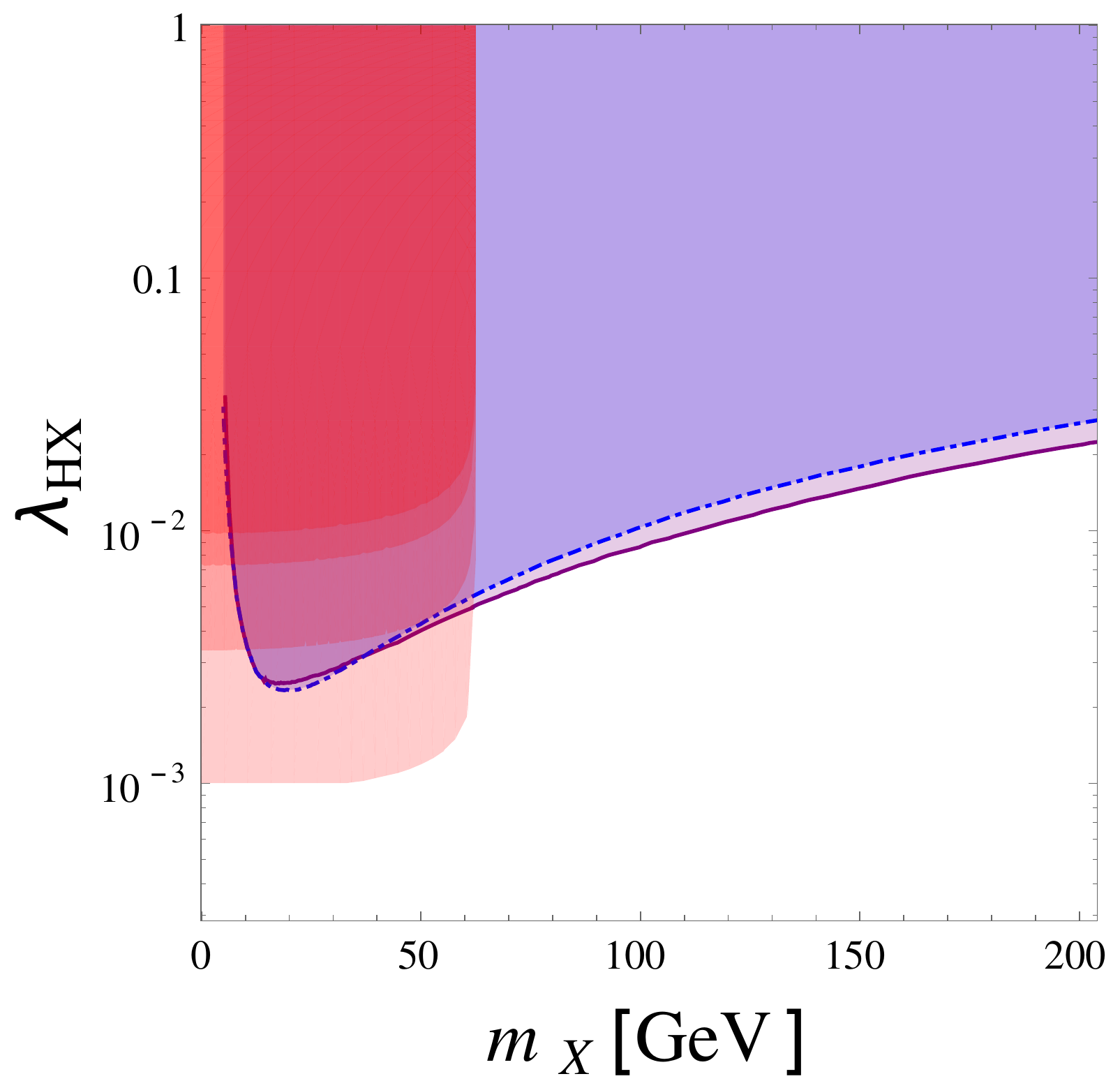} 
\caption{Exclusion regions in the ($m_X$,$\lambda_{HX}$)-plane.  The
  constraint from invisible Higgs decays is indicated in red.
 The continuous purple line correspond to 
  the recent limit reported by LUX~\cite{Akerib:2016vxi}
  from dark matter  searches while the dot-dashed blue 
  one indicates the current bound given by PandaX~\cite{Tan:2016zwf}.
  The different shades in red for the
  invisible decays define different contributions of
  $\Gamma^{\mathcal{D}}_{\text{Inv}}$ to $\mathcal{BR}_\text{Inv}$
  (see text).}
\label{fig:1Sconts}
\end{figure}

Encouraged by the above arguments concerning dark matter annihilation
and semi-annihilation processes and in view of the positive results of
Ref.~\cite{Belanger:2012zr}, here we take for granted that an adequate
relic abundance of the dark matter candidate particle can be ensured.
We focus, instead, on another most salient feature of our model,
namely, the presence of the invisible Higgs boson decays into
$Diracons$, i.e.
\begin{equation}
  \label{eq:inv}
h\to \mathcal{D D},  
\end{equation}
and its impact upon the dark matter detection prospects.
Such decays through $Diracon$ emission are the exact analogue of the
invisible Higgs decays by Majoron emission in models with Majorana
neutrinos~\cite{joshipura:1992hp}.
As long as the $h\to \mathcal{D D}$ coupling is non--zero, this Higgs
decay mode also contributes in the range $m_X< m_h /2$, that is, when
the Higgs decay into dark matter is kinematically allowed.
The current bound on the invisible Higgs decays is given by
$\mathcal{BR}_\text{Inv}\equiv
\frac{\Gamma_{\text{Inv}}}{\Gamma_{\text{Inv}}+\Gamma_{\text{Vis}}}<
17\%$~\cite{Bechtle:2014ewa}.  In this scenario, the invisible Higgs
decay width, $\Gamma_{\text{Inv}}$,  "always" has a contribution
coming from its decay into $Diracons$,
$\Gamma^{\mathcal{D}}_{\text{Inv}}\equiv \Gamma (h\to\mathcal{DD})$.
  As a result, for $m_X< m_h /2$, where $m_X$ is the dark matter mass,
  the invisible decays have two sources, the $h\to\mathcal{DD}$ and
  $h\to XX$, i.e.
  $\Gamma_{\text{Inv}}=\Gamma^{X}_{\text{Inv}}+\Gamma^{\mathcal{D}}_{\text{Inv}}$.
  The \sm Higgs is in general a combination of the doublet $H$ and the
  singlet $\sigma$, if we assume that the mixing between them is
  small, then
  $\Gamma_{\text{Vis}}$=$\Gamma^{SM}_{\text{Total}}=4.434$~MeV, so
  that the bound on the invisible width is
  $\Gamma_{\text{Inv}}$<0.908169~MeV.
  In this region there is a stronger constraint for the quartic
  coupling of the Higgs with the dark matter, as seen in
  Fig.~\ref{fig:1Sconts}. In this figure we display the constraints on
  $\lambda_{HX}$ from the invisible decays of the Higgs (red region)
  as well as from  the LUX~\cite{Akerib:2016vxi}
  and PandaX ~\cite{Tan:2016zwf} direct detection spin--independent cross
  section (purple and blue region, respectively).

The different shades in red in Fig.~\ref{fig:1Sconts} correspond to
different contributions of decays of Higgs into $Diracons$,
$\Gamma^{\mathcal{D}}_{\text{Inv}}$, the smaller the contribution of
$\Gamma^\mathcal{D}_{\text{Inv}}$, the darker the region.
For instance, the darkest red corresponds to the ``standard'' case
with $\Gamma^{\mathcal{D}}_{\text{Inv}}=0$, while the lightest one is
for $\Gamma^{\mathcal{D}}_{\text{Inv}}=0.9$~MeV.
As a result the region excluded by the invisible Higgs decays in the
($m_X$,$\lambda_{HX}$)--plane can be broader than the exclusion region
set by the LUX data for the mass range $m_X< m_h /2$. In other words,
the presence of the extra invisible decay channel into $Diracons$
effectively increases the sensitivities to spin--independent WIMP dark
matter searches below $m_h/2$.

\section{Summary}
\label{sec:summary}

In this letter we have proposed a low--scale mechanism for naturally
small Dirac neutrino masses generated only at the two--loop level. The
sector responsible for cosmological dark matter acts as messenger of
neutrino mass generation.  Both dark matter stabilitity and strict
lepton number conservation are ``symmetry protected''.
The presence of a global spontaneously broken U(1) symmetry leads to a
physical Goldstone boson, dubbed $Diracon$, that induces new invisible
Higgs decays detectable at LHC and other collider experiments.  The
coexistence of such decays with the Higgs to dark matter channel, if
kinematically allowed, leads to stronger sensitivities which we have
quantified using current constraints from the LHC. Detailed analysis
of the primordial WIMP dark matter density lies outside the scope of
the present letter and will be presented elsewhere.

\section*{Acknowledgements}

This work is supported by the Spanish grants FPA2011-2297,
FPA2014-58183-P, Multidark CSD2009-00064, SEV-2014-0398 (MINECO) and
PROMETEOII/2014/084 (GVA). EP is supported in part by PAPIIT IA101516
and PAPIIT IN111115. EP would like to thank the IFIC CSIC/UV for the
hospitality. We thank Roberto Lineros for discussions.

% % \bibliographystyle{T1}
% % \bibliography{merged_Valle,refs,newrefs,sco2loop.bib}

\end{document}